# P²-Gaussian: Prior-Driven and Projection-Guided Gaussian Splatting for Sparse-View Cone-Beam CT Reconstruction

Jinqiu Xia[1], Jianhui Ma[2], Bin Ouyang[1], Xiaoyan Huang[3], Yuan Xu[4*], Zun Piao[3*]

[1]Department of Radiation Oncology, The First Affiliated Hospital, Sun Yat-sen University, Guangzhou 510080, China; [2]Department of Radiation Oncology, Nanfang Hospital, Southern Medical University, Guangzhou 510515, China; [3]State Key Laboratory of Oncology in South China, Guangdong Key Laboratory of Nasopharyngeal Carcinoma Diagnosis and Therapy, Guangdong Provincial Clinical Research Center for Cancer, Sun Yat-sen University Cancer Center, Guangzhou 510060, PR China; [4]School of Biomedical Engineering, Southern Medical University, Guangzhou 510515, China

[*]Corresponding author: yuanxu@smu.edu.cn and piaozun@sysucc.org.cn

## Abstract

Sparse-view cone-beam computed tomography (CBCT) reconstruction reduces imaging dose and scan time, but severe angular undersampling makes the inverse problem highly ill-posed. Existing 3D Gaussian Splatting (3DGS) methods initialize Gaussian primitives from Feldkamp-Davis-Kress (FDK) reconstructions that are contaminated by streak artifacts. The limited angular coverage also leaves the Gaussian parameters underconstrained, producing needle-like artifacts. We propose P²-Gaussian, a prior-driven and projection-guided 3DGS framework. In this framework, a latent rectified flow model, Sparse2Prior, first restores the artifact-corrupted FDK volume into an anatomical prior in approximately 5 s. Gaussian primitives are then initialized from this prior in a tissue-adaptive manner, which yields a compact representation with 62%-65% fewer optimized primitives than random sampling. During optimization, virtual projections synthesized at intermediate angles between the measured projections are activated in rotation, while measured projections remain dominant in the objective function. Experiments on 5, 10 and 20 measured views showed that the proposed P²-Gaussian achieved the highest PSNR and SSIM among all methods. Compared with the optimization-based Gaussian baseline, our method increases PSNR by 5.98, 4.25 and 2.49 dB and accelerates reconstruction by 2.6–3.0 times. These results indicate that coupling a learned anatomical prior with measurement-constrained Gaussian optimization is effective for sparse-view CBCT reconstruction. Code is publicly available at: https://github.com/Ballbo5354/P-2-Gaussian.

## I. Introduction

Cone-beam computed tomography (CBCT)[1] has become an important three-dimensional imaging modality in modern medical imaging, with widespread applications in dental imaging[2,3], image-guided radiotherapy (IGRT)[4,5], and adaptive radiotherapy[6], owing to its rapid acquisition and high spatial resolution. During a typical course of radiotherapy, patients may undergo repeated CBCT scans for treatment positioning, with up to 30 CBCT-guided procedures reported in some clinical settings[7]. Repeated full-scan acquisitions increase cumulative radiation exposure and prolong acquisition time, thereby increasing the risk of patient motion and potentially compromising positioning accuracy. Therefore, enabling fast and accurate reconstruction from sparse view projections is important for reducing imaging dose and improving the efficiency of CBCT-guided radiotherapy.

However, severe projection undersampling makes CBCT reconstruction highly ill-posed. Analytical reconstruction methods, such as the Feldkamp Davis Kress (FDK) algorithm[8], often suffer from pronounced streak artifacts under sparse view acquisition. Iterative reconstruction methods, including the simultaneous algebraic reconstruction technique (SART)[9] and adaptive steepest descent - projection on convex sets (ASD-POCS)[10], can improve reconstruction quality by incorporating iterative data consistency and regularization. Nevertheless, their computational cost and slow convergence remain major limitations. These limitations motivate the development of more efficient reconstruction strategies for sparse-view CBCT.

Recent advances in deep learning have provided new opportunities for sparse-view CBCT reconstruction. Existing learning-based methods can generally be categorized into post-processing and direct reconstruction approaches. Post-processing methods, such as FBPConvNet[11] and FreeSeed[12], employ convolutional networks to suppress artifacts in analytically reconstructed images, while diffusion-based methods[13,14] further exploit learned generative priors to improve structural fidelity. Direct reconstruction methods instead learn mappings from projection measurements to volumetric images, evolving from early 2D encoder–3D decoder architectures[15] to geometry-aware feature backprojection and multi-view feature fusion strategies[16], with recent approaches such as DeepSparse incorporating large-scale pretraining and multi-scale feature integration[17]. Despite this progress, post-processing methods operate on an already reconstructed image and therefore cannot enforce consistency with the measured projections, whereas direct reconstruction methods learn a fixed feed-forward mapping that cannot be refined for an individual scan.

More recently, neural implicit representations have emerged as an alternative paradigm for sparse-view CBCT reconstruction. Neural Radiance Fields (NeRFs)[18] represent volumetric attenuation as continuous implicit functions and reconstruct volumes by optimizing neural fields. NAF[19], employs ray based volume rendering to reconstruct CBCT volumes and has demonstrated promising performance under sparse-view conditions. However, the computational cost of per ray optimization results in relatively slow reconstruction. DIF-Net[20] improves efficiency and quality in extremely sparse-view scenarios by introducing deep intensity fields and test time optimization. Nevertheless, neural implicit representations encode attenuation as continuous functions rather than explicit geometric primitives. This limits direct access to the reconstructed attenuation structure and makes targeted post-reconstruction refinement less straightforward.

3D Gaussian Splatting (3DGS)[21] offers an explicit and efficient alternative representation for CBCT reconstruction. Unlike the implicit representation of NeRF, 3DGS models volumetric scenes using explicit, differentiable anisotropic Gaussian primitives, enabling efficient optimization through splatting-based differentiable rendering. X-Gaussian[22] first extended 3DGS to X-ray imaging by introducing a radiative intensity response function and a differentiable radiative rasterization framework. $R^2$-Gaussian[23] rectifies the integration bias in splatting-based rendering, achieving reconstruction performance competitive with iterative methods while reducing optimization time by approximately 12-fold compared with NeRF-based approaches. DIF-Gaussian[24] adopts a feature-driven, feed-forward architecture. It extends DIF-Net by combining sparse-view projection features with spatially weighted features from a 3D Gaussian representation. The fused features are then used to predict voxel-wise attenuation coefficients, achieving competitive reconstruction performance with second-level inference.

Despite these advances, important challenges remain for 3DGS-based sparse-view CBCT reconstruction, particularly under severe projection undersampling. For optimization-based methods, such as $R^2$-Gaussian and its variants, reconstruction quality strongly depends on the reliability of the initial Gaussian representation. Under extremely sparse-view acquisition ($\leq 10$ views), the FDK volume is severely corrupted by undersampling artifacts, providing an unreliable initialization for subsequent Gaussian optimization and potentially biasing the recovery of anatomical structures. Meanwhile, the limited number of projection measurements provides insufficient angular information to constrain the large number of Gaussian parameters, making the optimization highly ill-posed and prone to inaccurate attenuation estimation, suboptimal convergence, and needle-like artifacts[25].

To address these limitations, we propose $P^2$-Gaussian, a prior-driven and projection-guided Gaussian Splatting framework for sparse-view CBCT reconstruction. Specifically, we propose Sparse2Prior, a latent rectified flow network that generates a high-quality prior volume from a sparse-view FDK volume. This prior volume is then used to obtain an anatomy-guided Gaussian initialization (AGI), which allocates Gaussian primitives according to tissue type and local gradient information. To further constrain the optimization under limited angular sampling, a virtual projection pool with epoch-wise rotation (VPPER) supplements the measured projections with virtual projections rendered from the prior at intermediate angles. By combining prior-driven initialization with projection supervision, $P^2$-Gaussian converges faster and reconstructs more accurately than existing Gaussian-based methods. The main contributions of this work are summarized as follows:

**1) A learned anatomical prior for sparse-view CBCT reconstruction**. We design a latent rectified flow model Sparse2Prior tailored for sparse-view FDK volume restoration, whose outputs serve as anatomically meaningful priors in our prior-driven reconstruction strategy. Notably, Sparse2Prior generates a prior in approximately 5 s per case.

**2) Anatomy-guided initialization and projection-guided optimization.** To exploit the prior at both the initialization and the optimization stage, we design AGI, which allocates Gaussian primitives according to tissue type and local gradient information and thereby reduces the number of primitives retained after densification by more than 60%, and VPPER , which precomputes a pool of virtual projections at the intermediate angles between the measured views and activates a small subset of the pool in each epoch, so that the measured projections remain dominant in the objective.

**3) Improved quality–efficiency trade-off.** By combining prior-driven initialization with projection-guided optimization, P²-Gaussian yields substantially improved reconstruction quality and faster convergence than existing Gaussian-based methods, particularly in severely sparse-view settings.

## II. Methods

### A. Preliminaries

#### A.1. 3D Gaussian Splatting for CBCT Reconstruction

3DGS represents a volumetric attenuation field using a set of explicit anisotropic Gaussian primitives. Following R²-Gaussian, the attenuation field is modeled as a collection of *M* anisotropic Gaussian primitives:

$$\mathbb{G}^3 = \{G_i^3\}_{i=1,\dots,M} \tag{2.1}$$

where each primitive is parameterized by a density coefficient $\rho_i$, center position $\boldsymbol{p}_i \in \mathbb{R}^3$, and covariance matrix$\Sigma_i \in \mathbb{R}^{3\times3}$. Accordingly, the attenuation coefficient at an arbitrary spatial location x is represented as:

$$\mu(\boldsymbol{x}) = \sum_{i=1}^{M} G_i^3\,(\boldsymbol{x} \mid \rho_i, \boldsymbol{p}_i, \boldsymbol{\Sigma}_i) = \sum_{i=1}^{M} \rho_i \exp\left(-\frac{1}{2}(\boldsymbol{x}-\boldsymbol{p}_i)^T \Sigma_i^{-1} (\boldsymbol{x}-\boldsymbol{p}_i)\right) \tag{2.2}$$

Unlike natural image rendering, X-ray image formation follows the Beer–Lambert law. Given an X-ray $\boldsymbol{r}$, the corresponding projection is obtained by integrating the attenuation field along its propagation path:

$$P(\boldsymbol{r}) = \int_{L(\boldsymbol{r})} \mu\,(\boldsymbol{x})\,d\boldsymbol{x} \tag{2.3}$$

For X-ray imaging, R²-Gaussian[23] removes the view-dependent spherical harmonics used in the original 3DGS formulation and introduces radiative Gaussian kernels to account for the integration bias during Gaussian splatting. The projected Gaussian density is corrected by:

$$\alpha_i = \sqrt{2\pi|\tilde{\Sigma}_i|/|\hat{\Sigma}_i|} \tag{2.4}$$

where $\tilde{\Sigma}_i \in \mathbb{R}^{3\times3}$ denotes the covariance transformed into the ray space and $\hat{\Sigma}_i \in \mathbb{R}^{2\times2}$ is obtained by projecting $\tilde{\Sigma}_i$ onto the detector plane. Consequently, the projection integral can be efficiently approximated by accumulating all projected Gaussian kernels:

$$P(\boldsymbol{r}) \approx \sum_{i=1}^{M} G_i^2\left(\hat{\boldsymbol{x}} \,\middle|\, \sqrt{\frac{2\pi|\tilde{\Sigma}_i|}{|\hat{\Sigma}_i|}}\rho_i, \hat{\boldsymbol{p}}_i, \hat{\Sigma}_i\right) \tag{2.5}$$

where $\hat{\boldsymbol{p}}_i = \phi(\boldsymbol{p}_i)$, $\hat{\Sigma}_i = \boldsymbol{J}_i \boldsymbol{W} \boldsymbol{\Sigma}_i \boldsymbol{W}^\top \boldsymbol{J}_i^\top$ ,with $\phi(.)$ denoting the projection transformation and $\boldsymbol{J}_i$ the local affine Jacobian.

However, existing Gaussian-based methods remain challenging under ultra-sparse-view conditions due to unreliable FDK initialization and limited angular constraints, motivating the use of prior information and additional projection guidance in the proposed P²-Gaussian.

**A.2. Latent Rectified Flow**

Diffusion models achieve high-quality medical image synthesis but require many iterative denoising steps, and operating on high-resolution 3D volume is computationally expensive and memory-intensive. MAISI[26] and 3D MedDiffusion[27] therefore adopt latent diffusion model (LDMs ), which encode a volume into a compact latent space with a variational autoencoder (VAE) and perform the generative process there.

Rectified Flow (RF) provides a more efficient formulation by learning a deterministic transport process between a source distribution $\pi_0$ and a target data distribution $\pi_1$. Specifically, let $z_0 \sim \pi_0$ and $z_1 \sim \pi_1$ denote samples from the source and target distributions, respectively. RF defines the intermediate state through linear interpolation:

$$z_t = tz_1 + (1-t)z_0, \quad t \in [0,1] \tag{2.6}$$

The corresponding target velocity is obtained as:

$$dz_t = v_t(z_t)dt \tag{2.7}$$

where $v_t(.)$ denotes the time-dependent velocity field. The training objective minimizes the discrepancy between the predicted velocity and the ground-truth displacement:

$$\mathcal{L}_{RF} = \int_0^1 \mathbb{E}_{z_0,z_1,t}\left[\| v_\theta(z_t,t) - (z_1 - z_0) \|^2\right]dt \tag{2.8}$$

After training, the source sample is transported toward the target distribution by solving the following deterministic ordinary differential equation (ODE)

$$\frac{dz_t}{dt} = v_\theta(z_t,t) t \in [0,1] \tag{2.9}$$

The approximately straight transport trajectories learned by RF enable efficient generation with substantially fewer function evaluations than conventional diffusion models. In this work, RF is formulated in the latent space and conditioned on sparse-view FDK reconstructions to generate prior CBCT volumes.

### B. Proposed P²-Gaussian framework

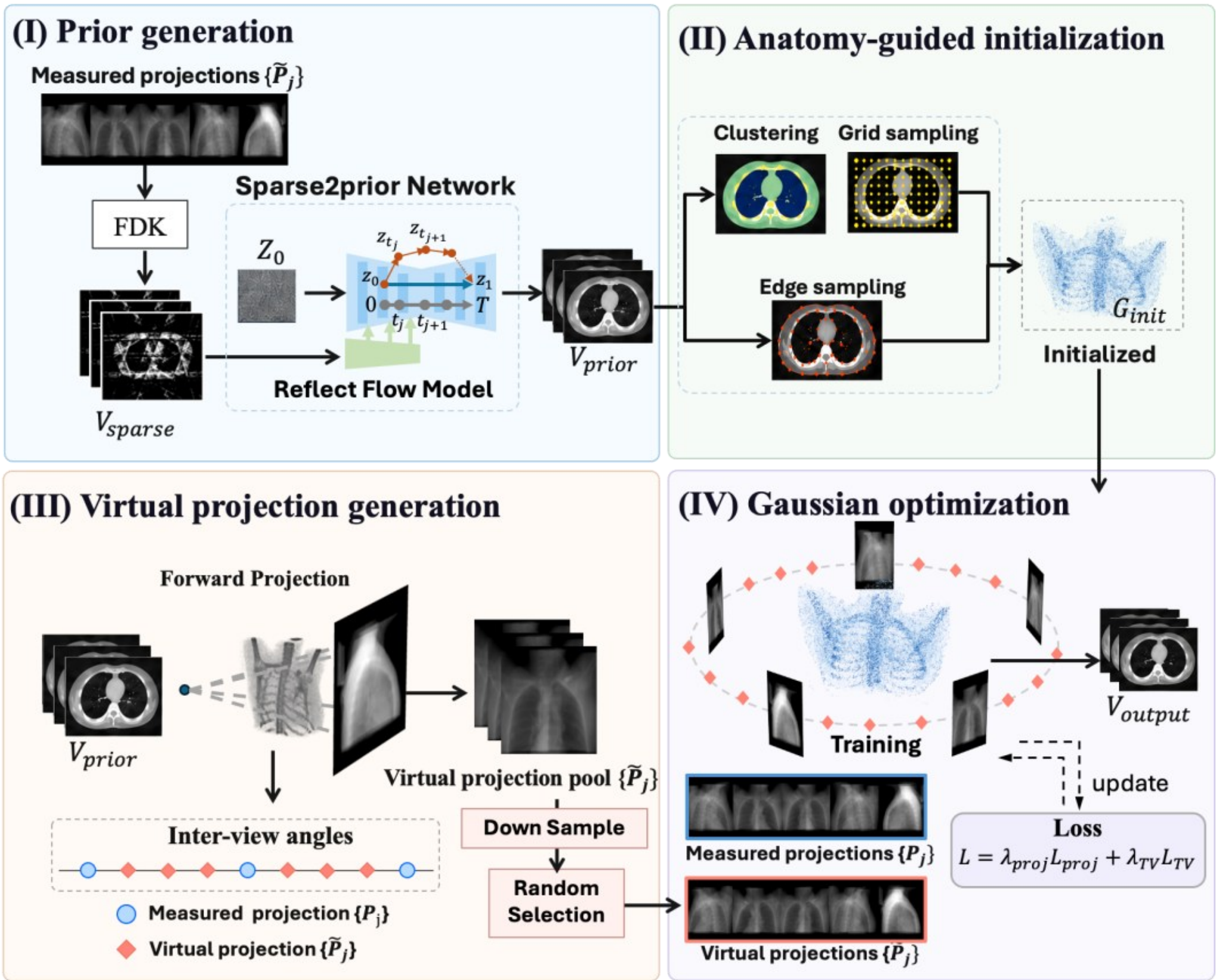

**Fig. 1.** The framework of proposed P²-Gaussian model.

The overall pipeline of the proposed P²-Gaussian framework is illustrated in Fig. 1, which consists of four stages:

**Stage I** generates an anatomical prior from sparse-view projections. Given sparse-view X-ray projections $\{P_j\}$, an FDK reconstruction $V_{\text{sparse}}$ is first obtained and then input to a ControlNet-conditioned RF model to generate a prior CBCT volume $V_{\text{prior}}$.

**Stage II** performs anatomy-guided Gaussian initialization. Instead of randomly sampling Gaussian primitives from the artifact-contaminated FDK reconstruction, the prior volume $V_{\text{prior}}$ is converted into an anatomically informed Gaussian point set $G_{\text{init}}$, which is used to initialize the subsequent 3DGS optimization.

**Stage III** generates virtual projections. Digitally reconstructed radiographs (DRRs) are synthesized from $V_{\text{prior}}$ at angles between the acquired views, producing a set of virtual projections $\{\tilde{P}_i\}$.

**Stage IV** optimizes the Gaussian representation using both measured and virtual projections. At each training iteration, a subset of virtual projections is randomly selected and combined with the measured projections for loss calculation. The measured projections are assigned the dominant contribution to the loss, while the virtual projections provide additional angular information.

These four stages allow P²-Gaussian to use the generated prior for Gaussian initialization and supplement the limited angular information with virtual projections, reducing the sensitivity of 3DGS optimization to FDK initialization and severe view sparsity.

### B.1. Rectified Flow Prior Generation

To obtain a reliable prior for sparse-view 3DGS reconstruction, we develop Sparse2Prior, a conditional Rectified Flow model for generating prior CBCT volumes. As shown in Fig. 2, it consists of a pretrained variational autoencoder (VAE), a pretrained 3D U-Net for Rectified Flow, and a trainable ControlNet. The VAE compresses a 3D CBCT volume into a four-channel latent space with a spatial compression ratio of four. During training, the VAE encoder and decoder, together with the 3D U-Net, are kept frozen using the weights released by MAISI-v2[28], and only the ControlNet is optimized.

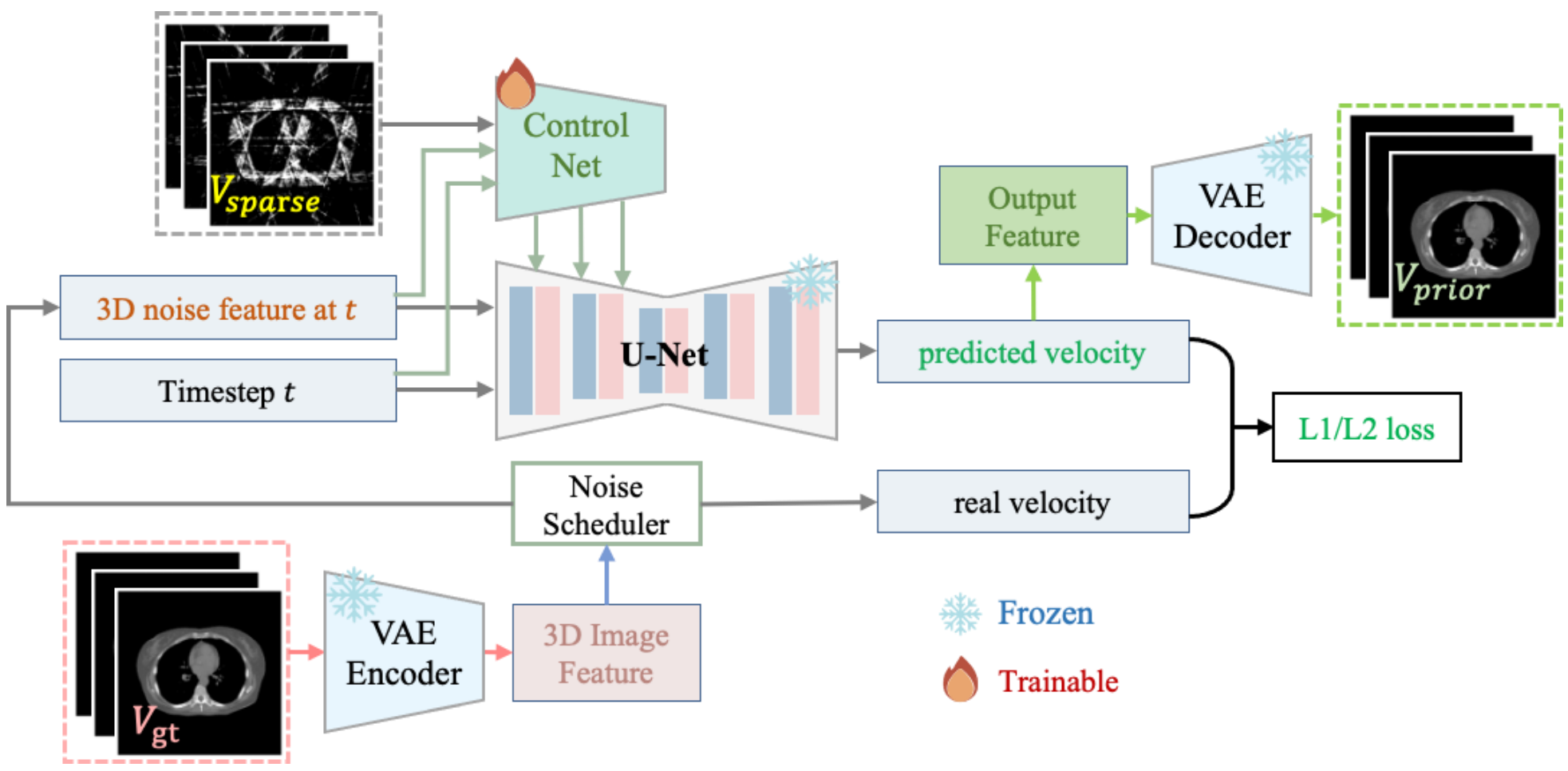

**Fig. 2.** The training framework of Sparse2Prior Network

Given a ground-truth CBCT volume $V_{gt} \in \mathbb{R}^{H\times W\times D}$, the frozen VAE encoder produces $z_1 \in \mathbb{R}^{\frac{H}{4}\times\frac{W}{4}\times\frac{D}{4}\times 4}$. The latent representations are precomputed before training and directly loaded during optimization. As show in part II.A.2, the source latent is sampled from $z_0 \sim \pi_0 = \mathcal{N}(0, I)$, while $z_1 \sim \pi_1$ denotes the latent representation of the ground-truth CBCT volume. ControlNet is implemented as a trainable copy of the pretrained 3D U-Net and uses the sparse-view FDK reconstruction $V_{sparse}$ as the conditional input. Given $V_{sparse}$, ControlNet extracts a conditional feature $c$ , which is injected into the frozen 3D U-Net through zero-initialized convolutions. Together with the noisy latent $z_t$ and flow time $t$, the conditional feature is used to estimate the conditioned velocity field:

$$\phi_{\text{Sparse2Prior}}(z_t, c) = v_\theta(z_t, t) + \mathcal{Z}(v_\theta(z_t + \mathcal{Z}(c; \Theta_{z1}); \Theta_c); \Theta_{z2}) \quad (2.10)$$

where $v_\theta(z_t, t)$ denotes the velocity field predicted by the frozen 3D U-Net, $\mathcal{Z}$ represents a zero-initialized $1 \times 1 \times 1$ convolution, and $\Theta_c, \Theta_{z1}, \Theta_{z2}$ are the trainable parameters of ControlNet.

The network is optimized using the conditional flow-matching objective:

$$\mathcal{L}_{\text{flow}} = \mathbb{E}_{z_0, z_1, t}\left[\| (z_1 - z_0) - \phi_{\text{Sparse2Prior}}(z_t, c) \|^2\right] \quad (2.11)$$

During inference, the $V_{sparse}$ is provided as the conditional input, and the latent state is initialized with Gaussian noise. The conditioned velocity field is integrated for 10 Rectified Flow sampling steps, after which the resulting latent representation is decoded by the frozen VAE decoder to obtain the prior CBCT volume $V_{prior}$. The $V_{prior}$ is then used to guide the anatomy-driven Gaussian initialization and the subsequent prior-guided 3DGS optimization.

### B.2. Anatomy-Guided Gaussian Initialization

To improve the initialization of Gaussian primitives under ultra-sparse-view CBCT reconstruction, we propose an Anatomy-Guided Gaussian Initialization (AGI) strategy. Instead of randomly sampling from artifact-contaminated FDK reconstructions, AGI generates anatomically informed Gaussian anchors from the prior volume $V_{prior}$ through three steps: tissue classification, tissue-adaptive grid sampling, and edge gradient augmentation, as illustrated in Fig.3.

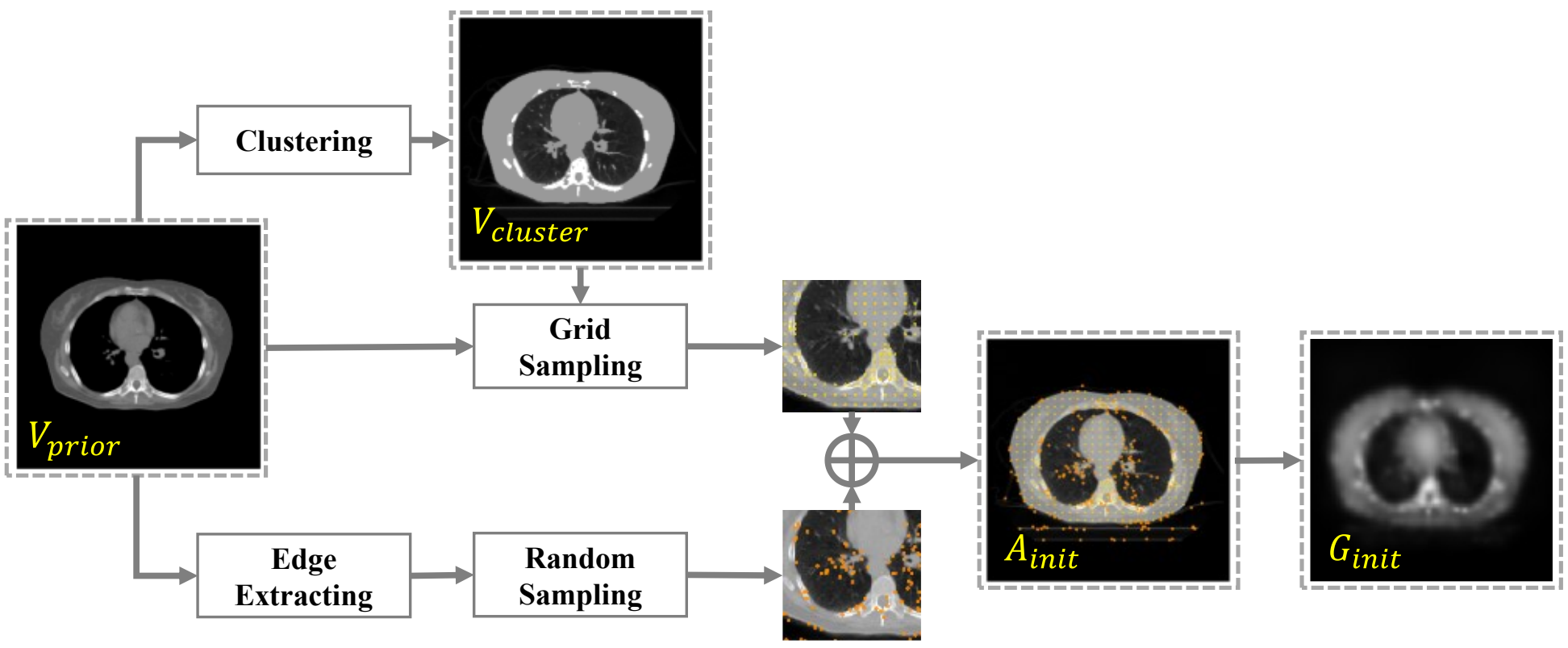

**Fig. 3.** The proposed Anatomy-Driven Gaussian Initialization

**Tissue classification.** Different anatomical tissues exhibit distinct spatial distributions and structural complexities. To exploit these characteristics, the prior volume $V_{prior}$ is first partitioned into three coarse tissue categories, including air, soft tissue, and bone. Specifically, K-means clustering is applied to the voxel intensities of $V_{prior}$, which provides a more robust tissue partition compared with manually predefined intensity thresholds.

Given the voxel intensity $V_{prior}(x_i)$at location $x_i$, the clustering objective is formulated as:

$$\{c_{air}, c_{\text{soft}}, c_{bone}\} = \underset{\{c_{air}, c_{\text{soft}}, c_{bone}\}}{\operatorname{argmin}} \sum_{i=1}^{X} \| V_{prior}(\boldsymbol{x}_i) - m_c \|^2 \tag{2.12}$$

where $m_c$ denotes the cluster center. The initial cluster centers are set as[29] :

$$\boldsymbol{m} = [0, 1.5 \times \mu_w, 2.6 \times \mu_w] \tag{2.13}$$

where $\mu_w$=0.192 represents the normalized linear attenuation coefficient of water. The clustering process produces three binary tissue masks:

$$M_{\text{air}}, M_{\text{soft}}, M_{\text{bone}} \in \{0,1\}^{H \times W \times D} \tag{2.14}$$

which provide tissue-specific spatial constraints for subsequent Gaussian anchor sampling.

**Tissue-Adaptive Grid Sampling.** Different tissue categories require different Gaussian sampling densities according to their anatomical characteristics. Air regions contain limited structural information and therefore require sparse sampling, whereas soft tissues occupy the majority of the imaging volume and require moderate sampling density. Bone structures contain abundant high-frequency anatomical details and are therefore sampled more densely.

Based on these observations, candidate Gaussian anchors are first generated using tissue-specific regular grids. Let $\sigma$ denote the base grid spacing. The sampling interval for each tissue category is defined as:

$$\Delta = \begin{cases} 2\sigma & \text{air (sparse)} \\ \sigma & \text{soft tissue (moderate)} \\ \max(1, \lfloor \sigma/2 \rfloor) & \text{bone (dense)} \end{cases} \tag{2.15}$$

For each tissue mask $M_c$, the corresponding candidate Gaussian anchor locations are defined as:

$$\boldsymbol{\mathcal{I}}_c = \{(i,j,k) \mid i,j,k \in \{0, \Delta, 2\Delta, \dots\}, M_c[i,j,k] = 1\} \tag{2.16}$$

Gaussian anchors are randomly sampled from each candidate set $\boldsymbol{\mathcal{I}}_c$. The resulting tissue-specific Gaussian anchor sets are denoted as: $A_{air}, A_{soft}, A_{bone}$.

**Edge Gradient Augmentation.** Although tissue-adaptive sampling preserves the global anatomical distribution, thin structures and tissue interfaces may still be insufficiently represented due to the regular grid-based sampling strategy. To further enhance local anatomical details, an edge-aware Gaussian anchor augmentation strategy is introduced.

Specifically, the gradient magnitude of $V_{prior}$ is first calculated to identify high-frequency anatomical regions. Voxels with gradient responses exceeding a predefined threshold are selected as candidate boundary anchors. Subsequently, $N_{edge}$ Gaussian anchors are randomly sampled from these candidates to form the boundary anchor set: $A_{edge}$.

Finally, the complete anatomical Gaussian anchor set is obtained by integrating tissue-aware and edge-aware anchors:

$$A_{init} = A_{air} \cup A_{soft} \cup A_{bone} \cup A_{edge} \tag{2.17}$$

where

$$|A_{init}| \leq N_{\mathrm{air}} + N_{\mathrm{soft}} + N_{\mathrm{bone}} + N_{\mathrm{edge}} \tag{2.18}$$

Each Gaussian anchor in $A_{\mathrm{init}}$ is subsequently converted into an initial Gaussian primitive by assigning its spatial position and the corresponding attenuation coefficient extracted from $V_{\mathrm{prior}}$ .The resulting initial Gaussian primitive set $G_{\mathrm{init}}$ is then used to initialize the subsequent 3DGS optimization.

### B.3. Virtual Projection Pool with Epoch-wise Rotation

Given the prior volume $V_{prior}$ produced in Stage I and the $M$ measured projection angles, Stage III supplies projection-domain supervision at angles where no measurement exists. We propose a VPPER, which first builds a pool of virtual projections at the angles left uncovered by the measured views, and then activates only a small subset of the pool in each optimization epoch. This provides complementary angular constraints while limiting the influence of any residual error in the prior volume.

### B.3.1 Virtual Projection Pool Construction

In this work the measured projections are acquired at uniformly spaced angles over the full $360°$ range, so the $M$ measured angles form a regular set:

$$\Theta_{\mathrm{meas}} = \{\theta_i\}_{i=1}^{M}, \qquad \theta_i = \theta_0 + (i-1)\,\frac{2\pi}{M} \tag{2.19}$$

which leaves an identical angular interval $2\pi/M$ between every pair of adjacent views. VPPER therefore refines this grid uniformly: let $N_{\mathrm{total}}$ denote the angular budget of the combined supervision set and $q = N_{\mathrm{total}}/M$ the refinement ratio, and define:

$$\Psi = \{\psi_n\}_{n=1}^{N_{\mathrm{total}}}, \qquad \psi_n = \theta_0 + (n-1)\,\frac{2\pi}{N_{\mathrm{total}}} \tag{2.20}$$

For integer $q$, the measured angles lie on this grid, $\Theta_{\mathrm{meas}} \subset \Psi$, and

$$\Psi = \Theta_{\mathrm{virt}} \cup \Theta_{\mathrm{meas}}, \qquad N_{\mathrm{pool}} = |\Theta_{\mathrm{virt}}| = N_{\mathrm{total}} - M \tag{2.21}$$

so the combined supervision set is uniformly spaced at $2\pi/N_{\mathrm{total}}$ for every value of $M$, and only the ratio of measured to virtual views changes with the acquisition. Each virtual projection is forward-projected from the prior volume and stored once,

$$\tilde{P}_j = \left\{\left(\theta_j,\ \hat{\mathcal{R}}_{\theta_j}^{\mathrm{virt}}(V_{\mathrm{prior}})\right)\right\}_{j=1}^{N_{\mathrm{pool}}}, \qquad \theta_j \in \Theta_{\mathrm{virt}} \tag{2.22}$$

where $\hat{\mathcal{R}}_{\theta}^{\mathrm{virt}}(\cdot)$ is the DRR projection operator at angle $\theta$. The pool is precomputed before the Gaussian optimization begins.

### B.3.2 Epoch-wise Rotation and Downsampled Supervision

To avoid excessive supervision from synthesized projections, VPPER activates only a small subset of $k$ virtual projections ($k \ll N_{\text{pool}}$) during each training epoch, so that an epoch comprises $M + k$ iterations. The subset is drawn from the least-recently-used members of the pool and spread over the uncovered angles by farthest-point sampling, which ensures that every virtual projection is activated before any is activated twice, so the full pool is traversed within $\lceil N_{\text{pool}}/k \rceil$ epochs.

Virtual projections are generated and rendered at a resolution downsampled by a factor $1/2$ relative to the measured projections. It reduces the per-iteration cost of the additional views. More importantly, downsampling low-pass filters the synthesized supervision, so the virtual views convey the coarse angular trend of the prior while fine detail is determined by the measured projections alone.

### B.4. 3DGS Optimization

Given the initialized primitive Gaussian set $G_{init}$, the measured projections $P_j$, and virtual projection pool $\widetilde{P}_j$, we iteratively optimize the Gaussian parameters using the differentiable projection framework of R²-Gaussian[30]. During optimization, both measured and virtual projections are incorporated into the projection-domain supervision. Measured projections provide reliable data-consistency constraints, whereas virtual projections supplement the missing angular information under sparse-view acquisition. Since virtual projections are generated from the prior image and may contain inherent prediction errors, their contribution is down-weighted to mitigate the propagation of prior-induced errors during Gaussian optimization.

The projection loss $\mathcal{L}_{\text{proj}}$ is defined as the $L_1$ distance between the rendered projection and the corresponding target projection. In addition, a three-dimensional total variation (3D-TV) term $\mathcal{L}_{\text{TV}}$ is imposed on randomly sampled sub-volumes from the voxelized 3DGS reconstruction to promote spatial smoothness and suppress local artifacts. The overall training objective is formulated as:

$$\mathcal{L}_{\text{total}} = \begin{cases} \mathcal{L}_{\text{proj}} + \lambda_{\text{TV}} \cdot \mathcal{L}_{\text{TV}}, & if\ view\ \in P_j, \\ \lambda_{virtual} \times \mathcal{L}_{\text{proj}} + \lambda_{\text{TV}} \cdot \mathcal{L}_{\text{TV}}, & if\ view\ \in \tilde{P}_j \end{cases} \quad (2.23)$$

where $\lambda_{\text{TV}}$ denotes the weight of the 3D-TV regularization. For measured projections $P_j$, the full loss is used to enforce consistency with the measured data. For virtual projections $\widetilde{P}_j$, the corresponding loss is scaled by $\lambda_{virtual}$, allowing the virtual projections to provide complementary angular constraints while reducing the influence of potential errors introduced by the prior reconstruction.

## III. Experiments

### A. Datasets and Preprocessing

In this study, two datasets were used to train and evaluate the Sparse2Prior Network: the publicly available AbdomenCT-1K dataset[31]and an internal clinical dataset consisting of 200 thoracic and abdominal CT cases. All cases were pooled and randomly shuffled before being split into training and test sets. A total of 50 cases were reserved for testing and designated as the Sparse2Prior test set, while the remaining cases were used for training. The performance of $P^2$-Gaussian and other reconstruction methods was evaluated on two additional independent datasets: the publicly available 4D-Lung dataset[32] and a separate internal clinical CT dataset. Following the data configuration of $R^2$-Gaussian, all CT volumes were resampled to 256×256×256. The TIGRE toolbox[33] was used to simulate X-ray projections and perform FDK reconstruction, with a source-to-detector distance (SDD) of 1500 mm and a source-to-axis distance (SAD) of 1000 mm. The simulated projections were generated with a detector resolution of 640 × 640. Sparse-view projections were uniformly sampled over the full 360° angular range. To evaluate reconstruction performance under different levels of view sparsity, experiments were conducted under 5, 10, and 20 projection views.

## B. Implementation Details

### B.1. Sparse2Prior Model

The Sparse2Prior model was trained on paired sparse-view FDK reconstructions and ground-truth CT volumes. The pretrained VAE and the 3D U-Net velocity backbone were kept frozen at the weights released with MAISI-v2[34], while only the ControlNet parameters were optimized using AdamW with a learning rate of $4 \times 10^{-5}$ and polynomial learning-rate decay for 100 epochs. Training was performed on a single NVIDIA RTX 3090 GPU. During inference, the RF ODE was solved with 10 steps to generate the prior CBCT volume from the sparse-view FDK reconstruction.

### B.2. Gaussian Initialization and Virtual Projection Configuration

For AGI, the sampling ratios for air, soft tissue, bone, and edge regions were empirically set to [0.05, 0.50, 0.25, 0.20], respectively. The grid sampling parameter $\sigma$ was set to 8, and the total number of initialized Gaussian primitives was fixed at 70,000. We evaluated reconstruction quality using different numbers of total projections to determine the size of the virtual projection pool. As shown in Fig. 4, the improvement became marginal beyond 60 views. Therefore, the total number of measured and virtual projections was fixed at 60, corresponding to 55, 50, and 40 virtual projections for the 5-, 10-, and 20-view settings, respectively. This configuration provides additional angular supervision while avoiding unnecessary computational overhead from excessive virtual projections.

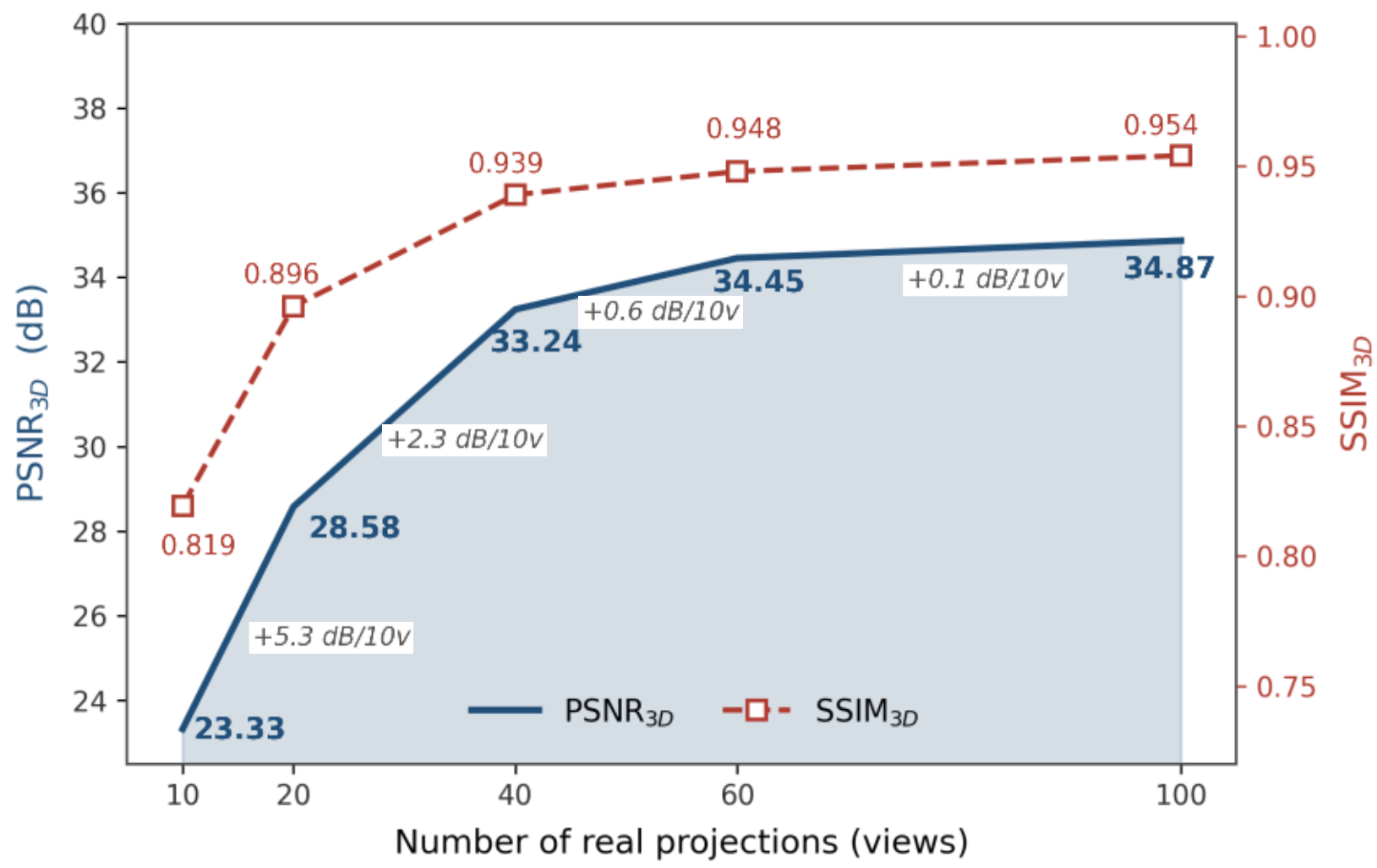

**Fig. 4.** Reconstruction quality versus the total number of projections after 8,000 optimization iterations, under uniformly distributed view configurations.

### B.3. 3DGS Optimization

The initialized Gaussian primitives were optimized for 10,000 iterations using Adam. The learning rates for Gaussian position, density, scale, and rotation were set to $1\times 10^{-4}$ , $5\times 10^{-3}$, $2\times 10^{-3}$ and $5\times 10^{-4}$, respectively, and were exponentially decayed to 10% of their initial values during optimization. Adaptive Gaussian densification was performed from iterations 500 to 5000. The acquired projections were used as the primary supervision, while the virtual projection loss was assigned a weight of $\lambda_{virtual} = 0.2$ in Eq. (2.23). All experiments were conducted on a single NVIDIA RTX 3090 GPU.

### C. Comparison Methods and Evaluation Metrics

We compared P²-Gaussian with representative reconstruction methods spanning conventional analytical and iterative reconstruction, implicit neural representation, and 3DGS. Specifically, FDK, ASD-POCS and SART were implemented using the TIGRE toolbox as conventional analytical and iterative baselines, respectively. NAF was included as a representative implicit neural reconstruction method, while R²-Gaussian was selected as a representative 3DGS-based approach. DIF-Gaussian was included as a representative feature-driven Gaussian reconstruction method. All methods were evaluated under the same projection geometry and sparse-view sampling conditions. Reconstruction quality was quantitatively assessed using peak signal-to-noise ratio (PSNR) and structural similarity index measure (SSIM). The average reconstruction time was recorded to evaluate the computational efficiency of each method.

## IV. Results and Analysis

### A. Qualitative Results

Fig. 5 compares the ground-truth (GT), sparse-view FDK reconstructions, and generated priors on the Sparse2Prior test set. Severe streak artifacts and structural distortion are observed in the FDK reconstructions under sparse-view conditions. The generated priors substantially reduce these artifacts and recover the main anatomical structures, showing good structural agreement with the GT images across the pelvic and thoracic regions. As the number of measured projections increases, the generated priors recover progressively finer anatomical details and become more consistent with the GT images. These results demonstrate that Sparse2Prior provides a more reliable prior volume than the artifact corrupted FDK reconstruction for subsequent Gaussian initialization.

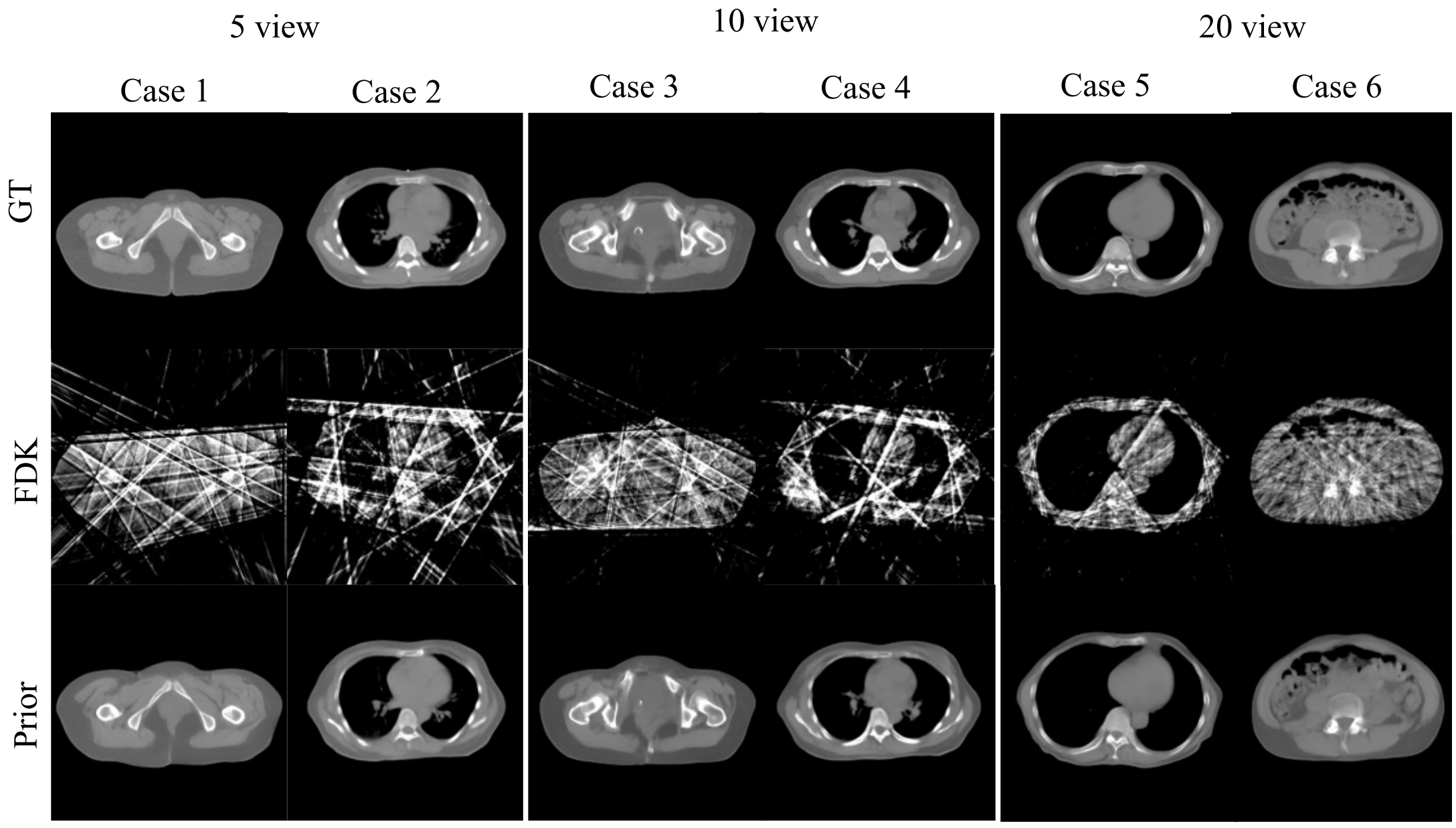

**Fig. 5.** Comparison of ground truth, FDK, and Sparse2Prior generated prior volumes under 5, 10, and 20 views settings.

Fig. 6 presents qualitative comparisons on a representative case from the 4D-Lung dataset under 5, 10, and 20 view settings. With only 5 views, the FDK reconstruction suffers from extensive streak artifacts throughout the field of view, while SART and ASD-POCS suppress some of the streak artifacts but introduce noticeable over-smoothing. NAF and $R^2$-Gaussian recover the overall thoracic structure but still exhibit structural distortions and loss of fine details. DIF-Gaussian provides a relatively clear reconstruction, but some anatomical boundaries and tissue details remain less consistent with the GT. In comparison, $P^2$-Gaussian preserves the overall anatomical structure more completely, with clearer lung boundaries, mediastinal structures, and chest wall contours, and shows better structural consistency with the GT. As the number of measured projections increases, all methods show improved reconstruction quality, although the degree of improvement varies across methods. $P^2$-Gaussian consistently preserves more complete anatomical structures and recovers finer details as the number of views increases from 5 to 20. The corresponding PSNR/SSIM values increase from 26.64 dB/0.8790 at 5 views to 28.47 dB/0.9046 at 10 views and 31.18 dB/0.9352 at 20 views.

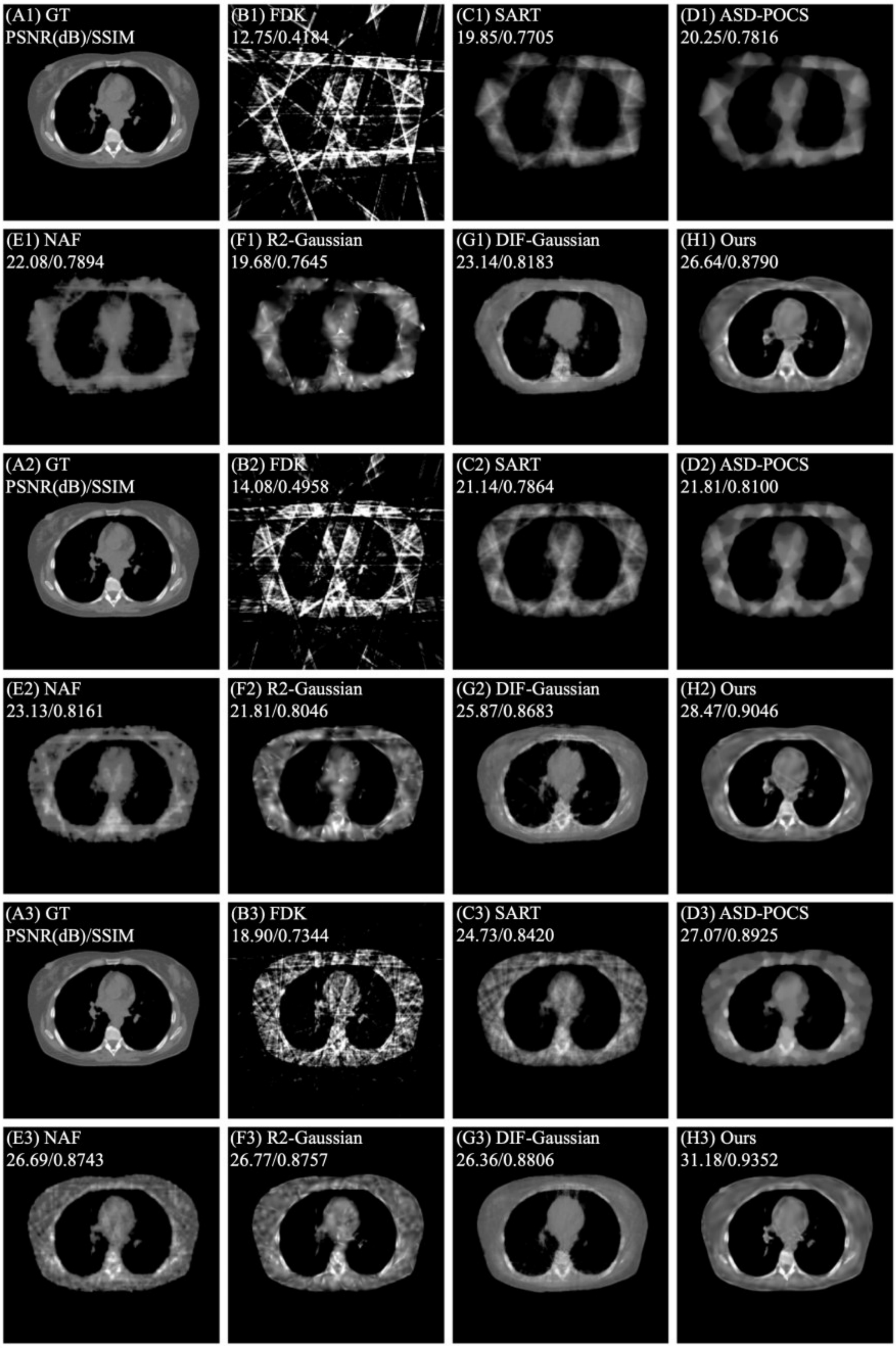

**Fig. 6.** Qualitative comparison of reconstructions for Case #101 from the 4D-Lung dataset. Window: [-400, 500] HU. A1–H1 Results under the 5-view setting; A2–H2 results under the 10-view setting; A3–H3 results under the 20-view setting.

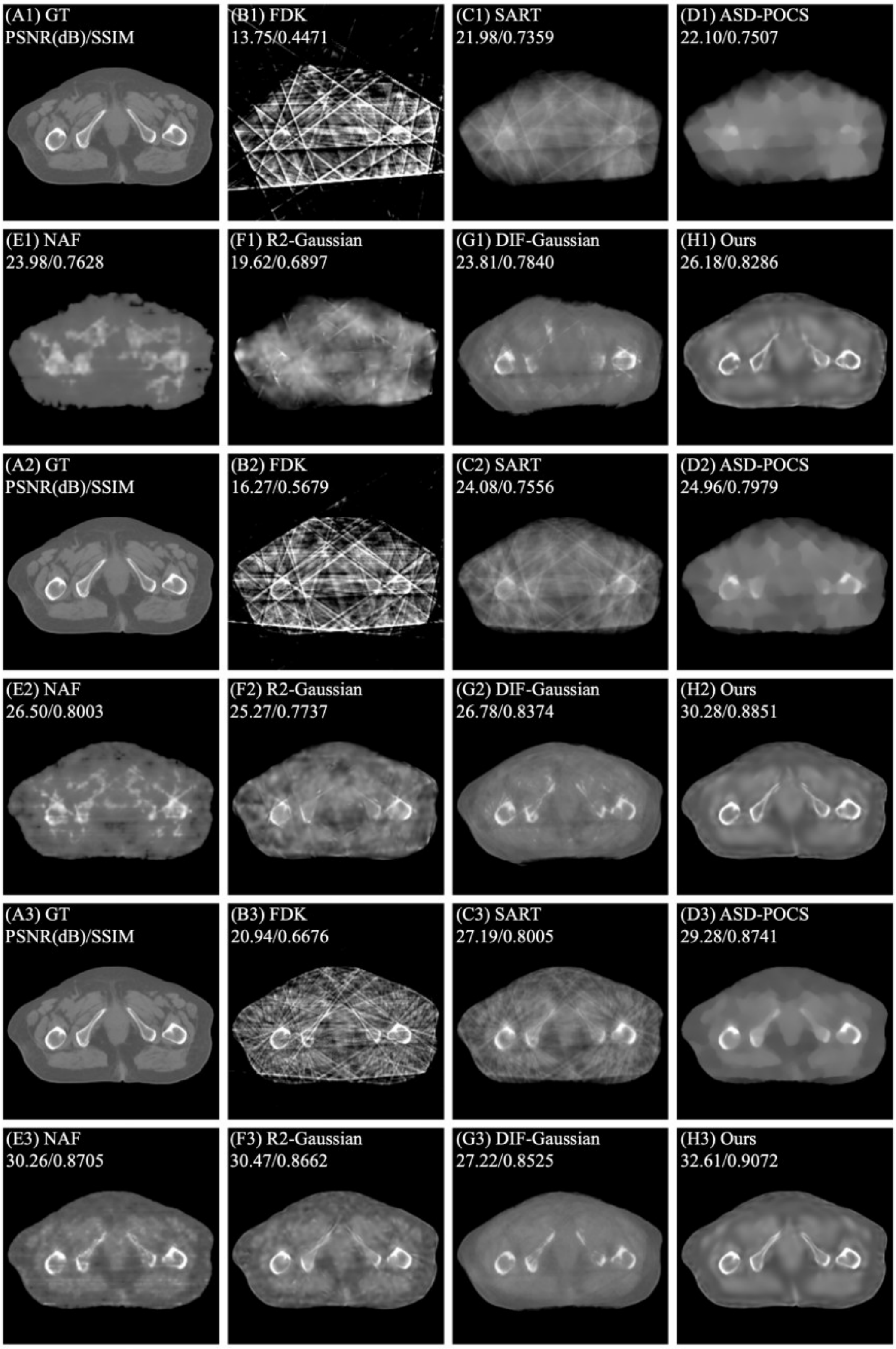

**Fig. 7.** Qualitative comparison of reconstructions for Case #66 from the internal dataset. Window: [-400, 500] HU. A1–H1 Results under the 5-view setting; A2–H2 results under the 10-view setting; A3–H3 results under the 20-view setting.

Fig. 7 shows the reconstruction results of an internal pelvic CBCT case under 5, 10, and 20 view settings. With only 5 views, FDK suffers from severe streak artifacts that obscure the pelvic structures, while SART and ASD-POCS still exhibit noticeable residual artifacts and over-smoothing. $R^2$-Gaussian and NAF improve the overall structural recovery but retain residual distortions and blurred high-frequency details. DIF-Gaussian provides better artifact suppression but fails to fully preserve fine osseous structures, resulting in less distinct cortical boundaries of the pelvic ring and femoral heads. In contrast, $P^2$-Gaussian produces sharper bone contours and more consistent soft-tissue structures, preserving both high-frequency skeletal details and low-contrast anatomical information. As the number of measured projections increases to 20 views, all methods show improved reconstruction quality, although residual streak artifacts and blurred cortical structures remain visible in the competing methods. $P^2$-Gaussian further improves the delineation of the pelvic ring and femoral heads and achieves a PSNR/SSIM of 32.61 dB/0.9072 at 20 views.

**B. Quantitative Results**

Table I quantitatively compares the FDK reconstructions with the prior volumes generated by Sparse2Prior under 5, 10, and 20 view settings. Sparse2Prior consistently achieves substantially higher PSNR and SSIM than FDK at all view numbers. At 5 views, the PSNR and SSIM improve from 14.36 ± 0.82 dB and 0.491 ± 0.057 to 26.85 ± 1.41 dB and 0.902 ± 0.042, respectively. The improvements remain evident at 10 and 20 views, with PSNR gains of 12.16 dB and 10.71 dB, respectively. These results demonstrate that Sparse2Prior can generate higher quality prior volumes than FDK, providing a more reliable initialization for subsequent Gaussian optimization.

Table I. Quantitative comparison of prior volumes generated by FDK and Sparse2Prior on the Sparse2Prior test datasets.

| Views | FDK | | Sparse2Prior | |
|---|---|---|---|---|
| | PSNR(dB) | SSIM | PSNR(dB) | SSIM |
| 5 views | 14.36±0.82 | 0.491±0.057 | 26.85±1.41 | 0.902±0.042 |
| 10 views | 16.79±0.85 | 0.621±0.063 | 28.95±1.34 | 0.926±0.031 |
| 20 views | 20.79±1.30 | 0.782±0.079 | 31.50±1.33 | 0.950±0.022 |

Table II summarizes the quantitative results under 5-, 10-, and 20-view settings. SART and ASD-POCS outperform FDK, while NAF achieves competitive reconstruction quality at a higher computational cost. DIF-Gaussian offers the fastest inference but yields lower PSNR and SSIM than $P^2$-Gaussian. $R^2$-Gaussian improves substantially with increasing view numbers, with its PSNR rising from 19.53 dB at 5 views to 27.83 dB at 20 views. The Sparse2Prior results provide a reference for the prior volume used to initialize $P^2$-Gaussian. After Gaussian optimization against the acquired projections, $P^2$-Gaussian further improves reconstruction quality, achieving the highest PSNR and SSIM among the evaluated reconstruction methods at all three view settings: 25.51 dB/0.847, 27.81 dB/0.889, and 30.32 dB/0.922 for 5, 10, and 20 views, respectively. Compared with $R^2$-Gaussian, it improves PSNR by 5.98, 4.25, and 2.49 dB and SSIM by 0.124, 0.101, and 0.056, respectively. The larger gains at lower view counts highlight the benefit of prior-guided reconstruction under severe view sparsity. Reconstruction takes 4 min 13 s, 4 min 25 s, and 4 min 39 s, respectively, including 5 s for prior generation, with only modest variation in runtime across view settings.

Table II. quantitative comparison on test datasets (4D-Lung and the internal clinical CT set). The best performance is shown in **bold**.

| Method | 5 views | | | 10 views | | | 20 views | | |
|---|---|---|---|---|---|---|---|---|---|
| | **PSNR** | **SSIM** | **Time** | **PSNR** | **SSIM** | **Time** | **PSNR** | **SSIM** | **Time** |
| FDK | 13.29 | 0.4660 | - | 15.73 | 0.598 | - | 19.57 | 0.721 | - |
| SART | 20.76 | 0.754 | 20s | 23.08 | 0.782 | 38s | 26.00 | 0.819 | 1m28s |
| ASD-POCS | 21.11 | 0.769 | 19s | 24.07 | 0.816 | 27s | 28.04 | 0.879 | 52s |
| NAF | 22.15 | 0.774 | 4m50s | 24.85 | 0.817 | 6m1s | 27.64 | 0.867 | 12m32s |
| $R^2$-Gaussian | 19.53 | 0.723 | 11m5s | 23.56 | 0.788 | 12m53s | 27.83 | 0.866 | 13m46s |
| DIF-Gaussian | 23.22 | 0.800 | **1.3s** | 25.99 | 0.852 | **1.5s** | 26.48 | 0.869 | **1.8s** |
| *†Sparse2Prior* | *25.40* | *0.831* | *5s* | *27.51* | *0.885* | *5s* | *29.66* | *0.913* | *5s* |
| $P^2$-Gaussian (*Ours*) | **25.51** | **0.847** | 4m13s* | **27.81** | **0.889** | 4m25s* | **30.32** | **0.922** | 4m39s* |

(*†Sparse2Prior is the prior-generation module of the proposed $P^2$-Gaussian, and is listed only as a reference for initialization quality; *Includes the 5 s required for prior volume generation by the Sparse2Prior Network.*)

Fig. 8 compares the SSIM and PSNR convergence curves of our method and $R^2$-Gaussian under 5, 10, and 20-view settings. Our method consistently achieves faster convergence and higher reconstruction quality across all sparse-view scenarios. Benefiting from the introduced prior information, the proposed method reaches higher PSNR values within fewer iterations and maintains a clear performance advantage over $R^2$-Gaussian during optimization.

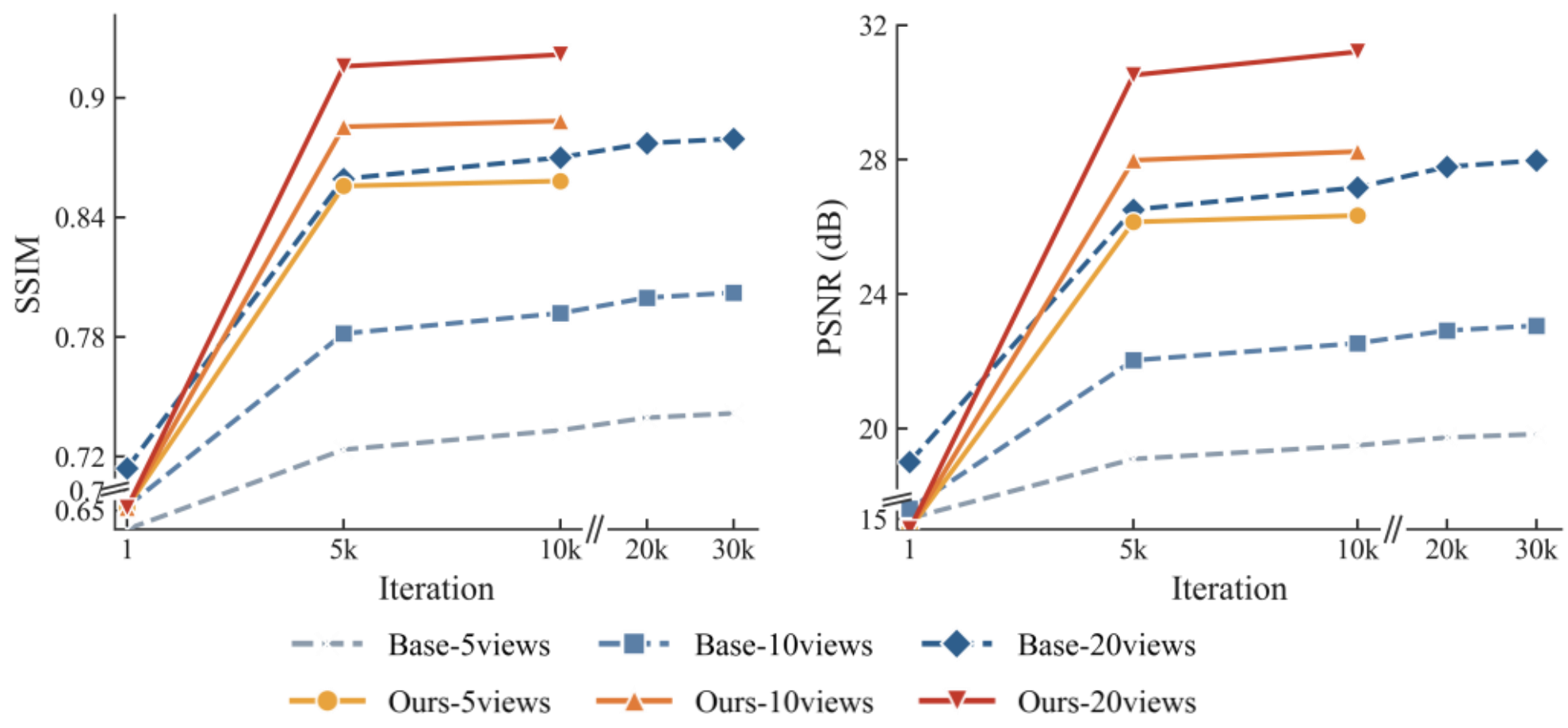

**Fig. 8.** Evolution of 3D PSNR over training iterations for the proposed $P^2$-Gaussian and the baseline $R^2$-Gaussian under the 5, 10, 20-view settings. The 3D PSNR is calculated between the reconstructed CBCT volumes and the ground-truth volumes.

### C. Ablation Study

To assess the contribution of the proposed components and their roles in sparse-view CBCT reconstruction, we conducted a series of ablation experiments on the 4D-Lung test dataset. Specifically, we examined the effects of three components of $P^2$-Gaussian: (i) the Sparse2Prior network, which provides a prior volume to replace the artifact-corrupted FDK reconstruction; (ii) AGI, which generates the initial Gaussian representation from the prior volume instead of conventional random sampling; and (iii) VPPER, which introduces complementary projection supervision to alleviate the limited angular coverage in sparse-view acquisition. Unless otherwise specified, all experiments were trained for 10k iterations on a single RTX 3090 GPU. The baseline uses the FDK reconstruction as the prior volume, conventional random Gaussian initialization, and no virtual projection supervision.

### C.1 Component Contribution

Table III. Ablation of individual components of P²-Gaussian under the 10-view setting. Prior: volume source for Gaussian initialization; Init.: Gaussian initialization type; VPPER: virtual projection pooling with epoch-wise rotation; Gau.: number of Gaussians after training. Best results in **bold**.

| Group | Prior | Init. | VPPER | PSNR | SSIM | PSNR-2D | SSIM-2D | Gau. |
|---|---|---|---|---|---|---|---|---|
| M1 | FDK | Random | | 22.43 | 0.787 | 36.90 | 0.984 | 129k |
| M2 | Sparse2Prior | Random | | 23.35 | 0.789 | 37.78 | 0.985 | 127k |
| M3 | Sparse2Prior | AGI | | 23.50 | 0.822 | **40.84** | 0.988 | 48k |
| M4 (P²-Gaussian) | **Sparse2Prior** | **AGI** | ✓ | **28.39** | **0.890** | 40.26 | **0.990** | **45k** |

Table III presents the progressive ablation results of P²-Gaussian under the 10-view setting, with each proposed component incrementally introduced into the baseline configuration. The baseline, using the FDK reconstruction as the prior volume, random Gaussian initialization, and no virtual projection supervision, achieves a PSNR/SSIM of 22.43 dB/0.787. Replacing the artifact-corrupted FDK reconstruction with the Sparse2Prior-generated prior increases the PSNR/SSIM to 23.35 dB/0.789, corresponding to a 0.92 dB gain in PSNR. This improvement suggests that the learned prior provides a more favorable initialization basis for subsequent Gaussian optimization by reducing the structural artifacts present in sparse-view FDK reconstruction. With the further introduction of AGI, the PSNR increases slightly to 23.50 dB, while SSIM improves from 0.789 to 0.822. A more pronounced improvement with PSNR-2D increasing from 37.78 dB to 40.84 dB. This indicates that anatomy-driven Gaussian initialization provides a more structured allocation of Gaussians according to the attenuation characteristics of different anatomical regions, improving structural and projection-domain consistency.

Finally, introducing VPPER produces the largest performance gain, increasing PSNR/SSIM to 28.39 dB/0.890, corresponding to improvements of 4.89 dB and 0.068, respectively, over the corresponding model without VPPER. The complete model also achieves the highest reconstruction quality with 45k Gaussians, compared with 129k for the baseline, indicating a substantially more compact Gaussian representation. Overall, the progressive improvements demonstrate the complementary roles of the prior volume, anatomy-driven initialization, and virtual projection supervision in sparse-view CBCT reconstruction.

Fig. 9 further shows the visual results of the four ablation settings under the 10-view condition. The baseline (M1) exhibits severe needle-like artifacts and fails to recover the bone contours accurately. Introducing the Sparse2Prior-generated prior (M2) improves the overall anatomical structure, but pronounced residual artifacts remain. With AGI (M3), the bone contours become more clearly defined and the needle-like artifacts are substantially suppressed, although residual artifacts are still present. The complete P²-Gaussian model (M4), with VPPER further incorporated, produces a markedly cleaner reconstruction, with further suppression of needle-like artifacts and more complete and consistent recovery of the bone structures. These qualitative results are consistent with the quantitative improvements reported in Table III.

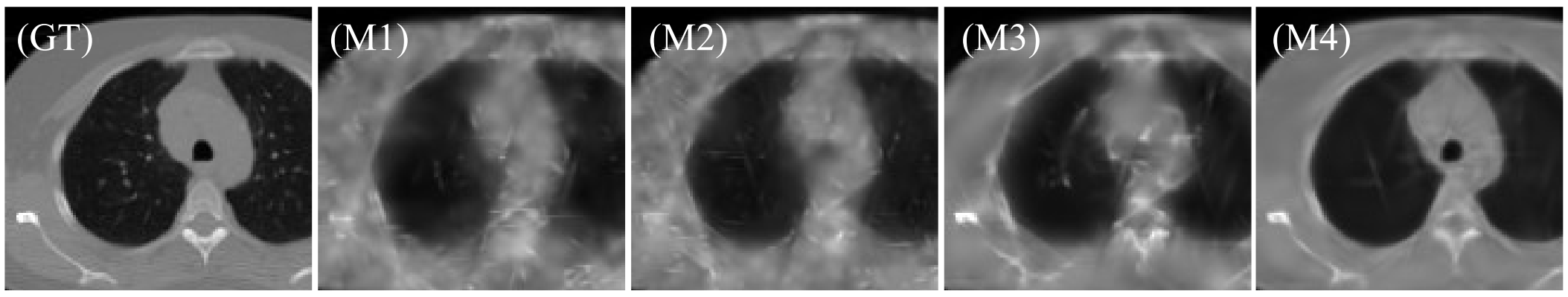

**Fig.9.** Qualitative comparison of the progressive ablation results of P²-Gaussian under the 10-view setting. M1–M4 correspond to the baseline, Sparse2Prior, Sparse2Prior + AGI, and the complete P²-Gaussian model with VPPER, respectively.

### C.2 AGI under Different Projection Numbers

Table IV evaluates the effectiveness of AGI under different projection numbers by comparing it with random Gaussian initialization while keeping the Sparse2Prior-generated prior and VPPER configuration fixed. AGI consistently improves reconstruction quality across all three view settings, yielding PSNR gains of 0.37, 0.49, and 1.15 dB under the 5-, 10-, and 20-view conditions, respectively. SSIM is also improved at all view settings, with gains of 0.016, 0.018, and 0.021, respectively. Meanwhile, AGI substantially reduces the number of optimized Gaussians from 107k–136k to 41k–50k, corresponding to an approximately 62%–63% reduction in representation size. Thus, AGI improves reconstruction fidelity while substantially reducing the number of Gaussian primitives, indicating a more compact and efficient representation under different levels of view sparsity.

Table IV: Effect of initialization strategy across projection numbers (fixed prior = Sparse2Prior, VPPER on). Δ: AGI vs. random; Gau.: number of Gaussians after training. Better results in bold.

| Views | PSNR (Rand.) | PSNR (AGI) | Δ | SSIM (Rand.) | SSIM (AGI) | Δ | Gau. (Rand.) | Gau. (AGI) |
|---|---|---|---|---|---|---|---|---|
| 5 | 27.12 | **27.49** | +0.37 | 0.864 | **0.880** | +0.016 | 107k | 41k |
| 10 | 27.90 | **28.39** | +0.49 | 0.873 | **0.890** | +0.018 | 117k | 45k |
| 20 | 29.48 | **30.49** | +1.15 | 0.893 | **0.914** | +0.021 | 136k | 50k |

### 3) VPPER under Different Projection Numbers

Table V further evaluates the effectiveness of VPPER under different projection numbers while keeping the Sparse2Prior-generated prior and AGI initialization fixed. VPPER consistently improves reconstruction quality across all three view settings, with the largest gain observed under the extremely sparse 5-view condition. At 5 views, VPPER increases PSNR/SSIM from 20.68 dB/0.766 to 27.49 dB/0.880corresponding to gains of 6.81dB and 0.114, respectively. At 10 and 20 views, the PSNR gains are 4.90 and 1.69 dB, while the SSIM gains are 0.068 and 0.014, respectively. The decreasing gains with increasing projection numbers indicate that VPPER is most beneficial under limited angular coverage. VPPER also has little effect on the final Gaussian count, which changes from 39k/48k/51k to 41k/45k/50k at 5, 10, and 20 views, respectively. Thus, VPPER improves reconstruction under sparse-view settings without substantially increasing the Gaussian representation size.

Table V: Effect of VPPER across projection numbers (fixed prior = Sparse2Prior, init = AGI). Δ: VPPER on vs. off; Gau.: number of Gaussians after training. Best results in bold.

| Views | PSNR (off) | PSNR (on) | Δ | SSIM (off) | SSIM (on) | Δ | Gau. (off) | Gau. (on) |
|---|---|---|---|---|---|---|---|---|
| 5 | 20.68 | **27.49** | +6.81 | 0.766 | **0.880** | +0.114 | 39k | 41k |
| 10 | 23.50 | **28.39** | +4.90 | 0.822 | **0.890** | +0.068 | 48k | 45k |
| 20 | 28.80 | **30.49** | +1.69 | 0.899 | **0.914** | +0.014 | 51k | 50k |

## V. Discussion & Conclusion

In this work, we propose P²-Gaussian, a prior-driven and projection-guided Gaussian Splatting framework for sparse-view CBCT reconstruction. P²-Gaussian is built upon the Gaussian reconstruction formulation of R²-Gaussian, while addressing the challenges introduced by severe view sparsity. R²-Gaussian provides an explicit Gaussian representation and an effective optimization framework for CBCT reconstruction, but it was not specifically designed for extremely sparse-view conditions. When only a few projections are available, the FDK reconstruction used for Gaussian initialization can contain severe streak artifacts and structural distortions, resulting in an unreliable initial Gaussian representation. At the same time, the limited angular information provides insufficient constraints for Gaussian optimization. These two factors make it difficult to recover accurate anatomical structures and can lead to suboptimal convergence under severe view sparsity.

The experimental results support this design. P²-Gaussian achieves its largest advantage over R²-Gaussian under severe view sparsity, improving PSNR by 5.98, 4.25, and 2.49 dB at 5, 10, and 20 views, respectively, with corresponding SSIM gains of 0.124, 0.101, and 0.056. The decreasing performance gap with increasing projection number is consistent with the ablation results, where VPPER provides larger gains under fewer views, confirming its role in supplementing limited angular coverage. AGI also reduces the number of optimized Gaussian primitives while maintaining or improving reconstruction quality, demonstrating that prior-guided initialization leads to a more compact and effective Gaussian representation.

Compared with other reconstruction approaches, P²-Gaussian combines high reconstruction quality with an explicit Gaussian representation that can be directly optimized using the measured projections. NAF relies on iterative optimization of an implicit neural representation and therefore requires longer reconstruction time. DIF-Gaussian adopts a feed-forward strategy, enabling fast inference with competitive reconstruction quality. Its end-to-end prediction differs from the explicit projection-driven optimization used in P²-Gaussian, providing less opportunity for subsequent refinement. Table II also includes the results generated by Sparse2Prior as a reference. Although Sparse2Prior achieves high image-domain fidelity, its generated prior is not constrained by the measured projections and therefore does not guarantee data consistency. P²-Gaussian uses this prior to initialize the Gaussian representation and provide complementary virtual projection supervision, while optimizing the Gaussian parameters against the acquired projections. By combining prior-derived anatomical information with measurement-driven Gaussian optimization, P²-Gaussian improves reconstruction fidelity while promoting consistency with the acquired data.

Despite these advantages, several limitations remain. P²-Gaussian still relies on iterative Gaussian refinement, which limits its reconstruction speed. Although the prior-guided initialization improves convergence, additional optimization is still required to obtain the final reconstruction quality, particularly as the number of projection views increases. In addition, the performance of Sparse2Prior may depend on the anatomical regions and acquisition conditions represented during training, which can affect its generalization to different clinical settings. Future work will focus on more efficient Gaussian optimization and improving the robustness and generalization of prior generation across anatomical regions and acquisition protocols.

Overall, P²-Gaussian demonstrates that combining anatomical prior construction with projection-guided Gaussian optimization is effective for sparse-view CBCT reconstruction. By providing a reliable initialization and complementary angular supervision while preserving fidelity to the acquired projections, the proposed framework improves reconstruction quality and convergence under severe view sparsity. These results support the use of prior-guided and measurement-constrained Gaussian optimization as a practical approach to sparse-view CBCT reconstruction.

## Reference

1. Nasseh I, Al-Rawi W. Cone Beam Computed Tomography. *Dent Clin*. 2018;62(3):361-391. doi:10.1016/j.cden.2018.03.002

2. Pauwels R, Araki K, Siewerdsen JH, Thongvigitmanee SS. Technical aspects of dental CBCT: state of the art. *Dentomaxillofacial Radiol*. 2015;44(1):20140224. doi:10.1259/dmfr.20140224

3. Alshomrani F. Cone-Beam Computed Tomography (CBCT)-Based Diagnosis of Dental Bone Defects. *Diagnostics*. 2024;14(13):1404. doi:10.3390/diagnostics14131404

4. Boda-Heggemann J, Lohr F, Wenz F, Flentje M, Guckenberger M. kV Cone-Beam CT-Based IGRT. *Strahlenther Onkol*. 2011;187(5):284-291. doi:10.1007/s00066-011-2236-4

5. Wilson LJ, Hadjipanteli A, Østergaard DE, et al. Cone beam CT dose optimisation: A review and expert consensus by the 2022 ESTRO Physics Workshop IGRT working group. *Radiother Oncol*. 2025;209:110958. doi:10.1016/j.radonc.2025.110958

6. Fischer J, Fischer LA, Bensberg J, et al. CBCT-based online adaptive radiotherapy of the prostate bed: first clinical experience and comparison to nonadaptive conventional IGRT. *Strahlenther Onkol*. 2025;201(8):767-778. doi:10.1007/s00066-024-02323-6

7. Lei W, Han L, Cao Z, et al. Nasopharyngeal cancer adaptive radiotherapy with CBCT-derived synthetic CT: deep learning-based auto-segmentation precision and dose calculation consistency on a C-Arm linac. *Radiat Oncol*. 2025;20(1):135. doi:10.1186/s13014-025-02715-7

8. Kak AC, Slaney M. *Principles of Computerized Tomographic Imaging: "This SIAM Edition Is an Unabridged Republication of the Work First Published by IEEE Press, New York, 1988."* Society for Industrial and Applied Mathematics (SIAM, 3600 Market Street, Floor 6, Philadelphia, PA 19104); 2001. doi:10.1137/1.9780898719277

9. Andersen AH, Kak AC. Simultaneous Algebraic Reconstruction Technique (SART): A superior implementation of the ART algorithm. *Ultrason Imaging*. 1984;6(1):81-94. doi:10.1016/0161-7346(84)90008-7

10. Sidky EY, Pan X. Image reconstruction in circular cone-beam computed tomography by constrained, total-variation minimization. *Phys Med Biol*. 2008;53(17):4777. doi:10.1088/0031-9155/53/17/021

11. Jin KH, McCann MT, Froustey E, Unser M. Deep Convolutional Neural Network for Inverse Problems in Imaging. *IEEE Trans Image Process*. 2017;26(9):4509-4522. doi:10.1109/TIP.2017.2713099

12. Ma C, Li Z, Zhang J, Zhang Y, Shan H. FreeSeed: Frequency-band-aware and Self-guided Network for Sparse-view CT Reconstruction. In: Vol 14229. 2023:250-259. doi:10.1007/978-3-031-43999-5_24

13. Pan S, Peng J, Gao Y, et al. CBCT Reconstruction Using Single X-Ray Projection With Cycle-Domain Geometry-Integrated Denoising Diffusion Probabilistic Models. *IEEE Trans Med Imaging*. 2025;44(7):2933-2947. doi:10.1109/TMI.2025.3556402

14. Xie J, Shao HC, Li Y, Zhang Y. Prior frequency guided diffusion model for limited angle (LA)-CBCT reconstruction. *Phys Med Biol*. 2024;69(13):135008. doi:10.1088/1361-6560/ad580d

15. Shen L, Zhao W, Xing L. Patient-specific reconstruction of volumetric computed tomography images from a single projection view via deep learning. *Nat Biomed Eng*. 2019;3(11):880-888. doi:10.1038/s41551-019-0466-4

16. Liu Z, Fang Y, Li C, et al. Geometry-Aware Attenuation Learning for Sparse-View CBCT Reconstruction. *IEEE Trans Med Imaging*. 2025;44(2):1083-1097. doi:10.1109/TMI.2024.3473970

17. Lin Y, Wang H, Chen J, Yang J, Guo J, Li X. DeepSparse: A Foundation Model for Sparse-View CBCT Reconstruction. *arXiv*. Preprint posted online May 5, 2025:arXiv:2505.02628. doi:10.48550/arXiv.2505.02628

18. Mildenhall B, Srinivasan PP, Tancik M, Barron JT, Ramamoorthi R, Ng R. NeRF: representing scenes as neural radiance fields for view synthesis. *Commun ACM*. 2022;65(1):99-106. doi:10.1145/3503250

19. Zhang Y, Li H. NAF: Neural Attenuation Fields for Sparse-View CBCT Reconstruction. In: Vol 13436. 2022:442-452. doi:10.1007/978-3-031-16446-0_42

20. Lin Y, Luo Z, Zhao W, Li X. Learning Deep Intensity Field for Extremely Sparse-View CBCT Reconstruction. *arXiv*. Preprint posted online August 31, 2023:arXiv:2303.06681. doi:10.48550/arXiv.2303.06681

21. Kerbl B, Kopanas G, Leimkuehler T, Drettakis G. 3D Gaussian Splatting for Real-Time Radiance Field Rendering. *ACM Trans Graph*. 2023;42(4):1-14. doi:10.1145/3592433

22. Cai Y, Liang Y, Wang J, et al. Radiative Gaussian Splatting for Efficient X-ray Novel View Synthesis. *arXiv*. Preprint posted online October 26, 2024:arXiv:2403.04116. doi:10.48550/arXiv.2403.04116

23. Zha R, Lin TJ, Cai Y, Cao J, Zhang Y, Li H. R$^2$-Gaussian: Rectifying Radiative Gaussian Splatting for Tomographic Reconstruction. *arXiv*. Preprint posted online October 27, 2024:arXiv:2405.20693. doi:10.48550/arXiv.2405.20693

24. Lin Y, Wang H, Chen J, Li X. Learning 3D Gaussians for Extremely Sparse-View Cone-Beam CT Reconstruction. *arXiv*. Preprint posted online July 7, 2024:arXiv:2407.01090. doi:10.48550/arXiv.2407.01090

25. Yuluo Y, Shen K, Ma Y, et al. GR-Gaussian: Graph-Based Radiative Gaussian Splatting for Sparse-View Tomographic Reconstruction. *IEEE Trans Comput Imaging*. 2026;12:1241-1253. doi:10.1109/TCI.2026.3701585

26. Guo P, Zhao C, Yang D, et al. MAISI: Medical AI for Synthetic Imaging. *arXiv*. Preprint posted online December 12, 2025:arXiv:2409.11169. doi:10.48550/arXiv.2409.11169

27. Wang H, Liu Z, Sun K, Wang X, Shen D, Cui Z. 3D MedDiffusion: A 3D Medical Latent Diffusion Model for Controllable and High-Quality Medical Image Generation. *IEEE Trans Med Imaging*. 2025;44(12):4960-4972. doi:10.1109/TMI.2025.3585372

28. Zhao C, Guo P, Yang D, et al. MAISI-v2: Accelerated 3D High-Resolution Medical Image Synthesis with Rectified Flow and Region-specific Contrastive Loss. *arXiv*. Preprint posted online August 7,

2025:arXiv:2508.05772. doi:10.48550/arXiv.2508.05772

29. Y. Zhang, H. Yu. Convolutional Neural Network Based Metal Artifact Reduction in X-Ray Computed Tomography. *IEEE Trans Med Imaging*. 2018;37(6):1370-1381. doi:10.1109/TMI.2018.2823083

30. Zha R, Lin TJ, Cai Y, Cao J, Zhang Y, Li H. R$^2$-Gaussian: Rectifying Radiative Gaussian Splatting for Tomographic Reconstruction. *arXiv*. Preprint posted online October 27, 2024:arXiv:2405.20693. doi:10.48550/arXiv.2405.20693

31. Ma J, Zhang Y, Gu S, et al. AbdomenCT-1K: Is Abdominal Organ Segmentation a Solved Problem? *IEEE Trans Pattern Anal Mach Intell*. 2022;44(10):6695-6714. doi:10.1109/TPAMI.2021.3100536

32. Hugo GD, Weiss E, Sleeman WC, et al. Data from 4D Lung Imaging of NSCLC Patients. Published online 2016. doi:10.7937/K9/TCIA.2016.ELN8YGLE

33. Biguri A, Dosanjh M, Hancock S, Soleimani M. TIGRE: a MATLAB-GPU toolbox for CBCT image reconstruction. *Biomed Phys Eng Express*. 2016;2(5):055010. doi:10.1088/2057-1976/2/5/055010

34. Zhao C, Guo P, Yang D, et al. Maisi-v2: Accelerated 3d high-resolution medical image synthesis with rectified flow and region-specific contrastive loss. In: *Proceedings of the AAAI Conference on Artificial Intelligence*. Vol 40. 2026:13088-13098.